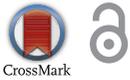

**APPLIED & INTERDISCIPLINARY MATHEMATICS | RESEARCH ARTICLE**

# A type D breakdown of the Navier Stokes equation in $d = 3$ spatial dimensions

Han Geurdes[1*]

**Abstract:** In this paper a type D breakdown of the Navier Stokes (NS) in $\mathbb{R}^d$ with $d = 3$, is demonstrated. The element of the breakdown also occurs in the Euler equation. We consider the fact that in $d = 2$, Ladyzhenskaya found a generalized type B solution. The discussion revolves around the notion, also found in quantum spin theory, that in $d = 2$ the behavior of a system can be quite different from the behavior in $d = 3$. Concerning applications, our resolution of the problem implies that e.g. $d = 3$ hydrology problems formulated as a NS equation can only be solved in computational approximation.

**Subjects:** Science; Earth Sciences; Mathematics & Statistics; Applied Mathematics; Applied Mechanics; Fluid Dynamics; Mathematical Physics

**Keywords:** type D breakdown; Clay Institute problem; Navier Stokes equation and Euler equation## 1. Introduction

The Navier Stokes (NS) equation is basic to many geophysical fluid mechanical problems. We mention here e.g. an application to groundwater flow (Masciopinto & Palmiotta, 2013) and coastal dynamics sediment transport giving dam breakdown (Morichon, Desombre, & Simian, 2013). There are however also many applications of the NS equation outside the field of geophysics. Basic physical aspects leading to the NS equation can be found in textbooks such as Prandtl's Essentials of Fluid Mechanics text (Oertel, 2004, p. 118 and further).

The problem of a possible exact solution of the NS equation in $\mathbb{R}^d$, with $d = 3$, is still open today. We cite here the transformation of the NS equation to a simpler form (Kozachok, 2013) and the work of Otelbayev (2013). Moreover, although Cafarelli c.s. produced an important inequality (Cafarelli, 1984) which is useful in the study of the NS equation, it is out of scope of the present paper. We also mention that for a finite time interval $[0, T)$ in $\mathbb{R}^3$, there is an exact solution that blows up when $T$ increases to untraceable large size (Fefferman, 2000). The exploding temporal behavior is a clue to a possible breakdown.

**ABOUT THE AUTHOR**

In our small engineering company we work on geo-data science problems. We work on hydrology on agriculture and we aim to begin to work on space weather problems.

**PUBLIC INTEREST STATEMENT**

In the present age, data analysis and mathematical modelling is becoming increasingly more important. Therefore it can be handy to know which mathematical problems have an exact solution and which problems do not. The Navier Stokes equation is used in many fields related to fluid dynamics. We mention for instance, acoustics, aerodynamics and space weather.

### *1.1. Aim of present work*

In the paper we aim for a breakdown of type D. The requirements for that type of breakdown in $d = 3$ can be found in Fefferman (2000). The key characteristic of the type D breakdown lies in $\left(\frac{Du}{Dt}\right)$. Because the Euler equation shares this form with the NS equation, the type D breakdown also occurs in the Euler equation. From a historical perspective, the NS equation is a viscous extension of the Euler equation Papanastasiou, Georgiou, and Alexandrou (2000) and Landau and Lifschitz (1987). This is also the starting point of the study of the linear non-stationary problem (Ladyzhenskaya, 1969, p. 81). Apparently, because $d = 2$ has an exact generalized solution, the addition of viscosity in fluid dynamics is most likely not causing a breakdown in $d = 2$. Moreover, although out of scope to the present paper, the cause of type D breakdown also resides in the Navier-Poisson equation as well (Papanastasiou et al., 2000, pp. 182–185).

The demonstration of type D breakdown starts with introducing the NS equation and the conditions to satisfy such a breakdown. Subsequently, particular velocity vectors are introduced and it is demonstrated that a breakdown force can match the initial velocity vectors. We find that, as required in a type D breakdown, no pressure function and velocity vector combination can be found in this case. An interesting extension of a possible breakdown in $d = 3$, namely the Oldroyd model in a viscoelastic NS equation Hynd (2013), will be discussed in a future study.

The breakdown velocity vectors are subsequently employed in a compound form and it is noted that, although the sum of the compound form is approximative, any selected point in time can be associated to a velocity breakdown equation. The reason is the fact that between two *different* points on the real axis, one has the continuum. In the limit (which only can be approximated) the compound form is exact.

The criterion for force is a force vector function which is independent of time and is incidentally also divergence free.

In $d = 2$ we have a generalized solution (Ladyzhenskaya, 1969). How does this relate to a type D breakdown in $d = 3$. In an analogy, we may note a case outside of continuum mechanics. In anyon spin statistics (Wilckzek, 1982), the reduction of spatial dimensions from $d = 3$ to $d = 2$ gives a profound change in the spin behavior of particles. In Fefferman (2000) the differences between a $d = 2$ and $d = 3$ NS equation are briefly mentioned.

Finally, as a practical consequence of the demonstrated type D breakdown, only a weak solution or an approximative numerical solution is allowed in the use of the incompressible viscous NS equation in $d = 3$. We mention fields such as geo-hydrology, hydro-meteorology, aerodynamics and space weather. In the latter case, for instance, ion drift studies relate the NS equation to electromagnetic phenomena (Shukla, 1982).

### *1.2. The requirements for the NS equation problem*

In the NS equation the velocity vector, $u$, $\{u_n\}_{n=1}^3$, is matched with a simultaneous solution for scalar pressure $p(x, t) \in C^N\left(\mathbb{R}^3 \times [0, \infty)\right)$ given the force $f_n(x, t)$ with $n = 1, 2, 3$. Here, $N \in \mathbb{N}$ and we can have $N \to \infty$, with $\infty$ meaning here, $N$ is untraceable large. In the breakdown we use $N$ untraceable large and growing if necessary, but finite, in order to satisfy the requirements of failure of solution. A possible equation for pressure, $p$ and the selected force, $f$, observe the required characteristics given in equations (8) and (11) of Fefferman (2000). Generally we have for the $n$-th element $u_n = u_n(x, t)$, ($n = 1, 2, 3$), with, $x = (x_1, x_2, x_3) \in \mathbb{R}^3$ of the velocity, the NS equation

$$\frac{\partial u_n}{\partial t} + (u \cdot \nabla)u_n - \nu \nabla^2 u_n = f_n - \frac{\partial p}{\partial x_n}, \quad (1.1)$$

with $\nu > 0$ and $(u \cdot \nabla) = \sum_{j=1}^{3} u_j \frac{\partial}{\partial x_j}$. It is noted that $u_n(x,t) \in C^N(\mathbb{R}^3 \times [0,\infty))$. The associated NS operator, related to (1.1), is subsequently defined by

$$\mathcal{N}_u = \left(\frac{\partial}{\partial t} + (u \cdot \nabla)\right) - \nu \nabla^2 = \frac{D}{Dt} - \nu \nabla^2. \quad (1.2)$$

The type-D resolution is formulated in Fefferman (2000) as follows: "… Take $\nu > 0$ (in (1.1)), and the space $\mathbb{R}^d$ with $d = 3$. Then there exists a smooth, divergence-free vector field $u^0(x)$ on $\mathbb{R}^3$ and a smooth $f(x,t)$ on $\mathbb{R}^3 \times [0,\infty)$, satisfying (Fefferman, 2000): (8), (9), i.e.

$$u^0(x + e_j) = u^0(x), \quad f(x + e_j, t) = f(x, t) \quad (1.3)$$

and

$$|\partial_x^\alpha \partial_t^m f(x,t)| \leq C_{\alpha,m,K}(1 + |t|)^{-K}, \quad (1.4)$$

for which there exist no solutions $(p, u)$ of, in Fefferman (2000): (1), (2), (3), (10), (11) on $\mathbb{R}^3 \times [0, \infty)$ …". The $\alpha$ is a $d = 3$ vector $\alpha = (\alpha_1, \alpha_2, \alpha_3) \in \mathbb{N}^3$ and $m \in \mathbb{N}$. In the present paper we will see that one can select $(u^0, f)$ such that it is not possible to find a $(p, u)$ in $d = 3$. For completeness, Fefferman's (1) is our (). Fefferman's (2) is $\nabla \cdot u = 0$. The required breakdown of (3) and (10) will be made clear in the text below. We interpret Fefferman's "… there exists no solution … $(p, u)$ … of …(11)" as having, at its least, the possibility of $(p, u) \in C^N$ with $N \in \mathbb{N}$ and $1 < N \sim$ large on $\mathbb{R}^3 \times [0, \infty)$. We note here that $\infty$ means untraceable large. Fefferman's (11) (Fefferman, 2000) reads, $p \in C^\infty(\mathbb{R}^3 \times [0, \infty))$ and for $n = 1, 2, 3$, we also have, $u_n \in C^\infty(\mathbb{R}^3 \times [0, \infty))$. $C^\infty$ is replaced here with $C^N$ and $N$ untraceable large.

## 2. The construction

### 2.1. Preliminaries

In the first place let us select any function $u_k^0(x) \in C^N(\mathbb{R}^3)$, $1 < N \sim$ large and, $k = 1, 2, 3$ with the following characteristics. The $u^0(x)$ vector in $\mathbb{R}^3$, must be divergence free. Moreover, according to Equation (8) of Fefferman (2000) we must observe spatial periodicity, (1.3) for $e_j$ unit vectors, $(e_j)_m = \delta_{j,m}$, with, $j, m = 1, 2, 3$. So the set of vector functions,

$$\mathcal{U} = \{u^0 \mid \nabla \cdot u^0(x) = 0, \ u^0(x + e_j) = u^0(x), \ u_k^0(x) \in C^N(\mathbb{R}^3)\}, \quad (2.1)$$

can be defined. Of course we may select any function from $\mathcal{U}$. Moreover, requirements for the force vector $f$ in (1.1) are captured in the set

$$\mathcal{G} = \left\{f_k \in C^N \mid f(x + e_j, t) = f(x, t), |\partial_x^\alpha \partial_t^m f(x,t)| \leq C_{\alpha,m,K}(1 + |t|)^{-K}\right\}, \quad (2.2)$$

with $j, k = 1, 2, 3$. Note, that $N$ in both (2.1) and (2.2) is in $\mathbb{N}$ with $1 < N$. Again, $\alpha = (\alpha_1, \alpha_2, \alpha_3)$ with the $\alpha_k$ together with $m$, finite elements of $\mathbb{N}$. So, $f \in \mathcal{G}$ where we must see arbitrary $\alpha \in \mathbb{N}^3, m \in \mathbb{N}$, both finite, and $K \in \mathbb{R}$. In addition, $\partial_n = \partial_{x_n} = \frac{\partial}{\partial x_n}$ and $\partial_t = \frac{\partial}{\partial t}$.

### 2.2. Initial definitions
Subsequently, let us define a vector function $\xi^h = \xi^h(x,t)$ with, for all, $h \in [0, \infty)$,

$$\xi^h(x,t) = x - (t-h)(x + u^0(x)). \tag{2.3}$$

For $t = h$, we have, $\xi^h(x,h) = x$. In addition, let us define the function $v^h(x,t)$ with

$$v^h(x,t) = c + x + u^0\left(\xi^h(x,t)\right). \tag{2.4}$$

The $c$ is a constant vector ($c \neq 0$), independent of $x \in \mathbb{R}^3, t \in \mathbb{R}, t \geq 0$ and $h$.

### 2.3. Requirements (2) (3) and (10) (Fefferman, 2000)
Equation (3) of Fefferman (2000) does not apply as required in a type D breakdown, i.e. we don't have Fefferman's $u(x,0) = u^0(x)$, because in our case from (2.4), $v^h(x,0) = c + x + u^0\left(\xi^h(x,0)\right)$.

From Equation (2.3) it easily follows that, as required in a type D breakdown looking at (3) in Fefferman (2000), for $h = 0$ we see $v^{(h=0)}(x,0) = c + x + u^0(x)$. This implies that $v^{(h=0)}(0,0) = c + u^0(0)$. Hence, for $h = 0$, $v^h(x,0) \neq u^0(x)$ as required. If $h \neq 0$ then

$$v^h(x,0) = c + x + u^0[x + h(x + u^0(x))] \neq u^0(x). \tag{2.5}$$

Note from (2.3) $\xi^h(x,0) = x - (0-h)(x + u^0(x))$.

In addition it follows from (2.4) that $\nabla \cdot v^h(x,t)|_{t=h} = \nabla \cdot x + \nabla \cdot u^0(x) = 3 \neq 0$. In this case the restriction $t = h$ is sufficient for requirement (2) of Fefferman (2000). More generally, however, we may write for the div of $v^h$, i.e. $\nabla \cdot v^h(x,t) = 3 + \nabla \cdot u^0[\xi^h(x,t)]$, leading to

$$\nabla \cdot v^h(x,t) = 3 - (t-h) \sum_{j=1}^{3} \sum_{k=1}^{3} \left(\frac{\partial u_k^0(x)}{\partial x_j}\right)\left(\frac{\partial u_j^0(\xi^h)}{\partial \xi_k^h}\right) \tag{2.6}$$

which is not identically zero. The $\{\xi_k^h\}_{k=1}^{3}$ are the entries of the vector $\xi^h \in \mathbb{R}^3$ defined in (2.3). Note, $\left(\frac{\partial \xi_j^h(x,t)}{\partial x_k}\right) = \delta_{j,k}(1 - (t-h)) - (t-h)\left(\frac{\partial u_j^0(x)}{\partial x_k}\right)$ and so in (2.6) there is the vanishing double sum of the divergence free $u^0$

$$(1 - t + h) \sum_{j=1}^{3} \sum_{k=1}^{3} \delta_{k,j} \frac{\partial u_j^0(\xi^h)}{\partial \xi_k^h} = 0.$$

$\delta_{k,j}$ is the Kronecker delta. Looking again at (2.6) and at (2.5), we may conclude that in our case, Equation (2) of Fefferman (2000) breaks down.

Moreover, $e_j$ shifts do not hold for $v^h$,

$$v^h(x + e_j, t) \neq v^h(x,t), \tag{2.7}$$

for all $x \in \mathbb{R}^3$. This is obviously true for $t \neq h$ looking at definition (2.3). It also holds, because of the linear $x$ term in (2.3), for $t = h$. In addition, noting that for $u^0 \in \mathcal{V}$ it follows e.g. $u^0(x + 2e_j) = u^0(x + e_j) = u^0(x)$. Hence, it can be observed from (2.3), that

$$\xi^h(x + e_j, t) = x + e_j - (t - h)(x + e_j + u^0(x)) = \xi^h(x, t) + (1 - (t - h))e_j.$$

For $u^0 \in \mathcal{U}$, the effect of $\xi^h(x + e_j, t) \neq \xi^h(x, t)$ in the $u^0(\xi^h)$ of (2.4) occurs if $t - h \notin \mathbb{Z}$ with $\mathbb{Z}$ the integer positive or negative numbers including zero. Hence, because both the linear $x$ in (2.4) *and* the definition of $\xi^h$ in (2.3), the $v^h$ behave as required in a D breakdown looking at (10) of Fefferman (2000). To wrap this up, we don't have Fefferman's

$$u(x, t) = u(x + e_j, t) \tag{2.8}$$

for our $v^h(x, t)$ and $j = 1, 2, 3$, as required in a D breakdown.

### 2.4. Differentiation to match terms in $\mathcal{N}_{v^h}$ i.e. requirement (1) (Fefferman, 2000)

In the first place let us differentiate $\xi^h$ to $t$. This implies

$$\frac{\partial \xi^h}{\partial t}(x, t) = -(x + u^0(x)). \tag{2.9}$$

So, if $v^h$ is differentiated to $t$ we have

$$\frac{\partial v^h(x, t)}{\partial t} = \left(\frac{\partial \xi^h}{\partial t} \cdot \nabla_{\xi^h}\right) u^0(\xi^h). \tag{2.10}$$

And,

$$\left(\frac{\partial \xi^h}{\partial t} \cdot \nabla_{\xi^h}\right) = \sum_{k=1}^{3} \frac{\partial \xi_k^h}{\partial t} \frac{\partial}{\partial \xi_k^h}.$$

Combining equations (2.9) and (2.10) gives

$$\frac{\partial v^h(x, t)}{\partial t} = -\left(\left(x + u^0(x)\right) \cdot \nabla_{\xi^h}\right) u^0(\xi^h). \tag{2.11}$$

Hence,

$$\frac{\partial v^h(x, t)}{\partial t}\bigg|_{t=h} = -\left(\left(x + u^0(x)\right) \cdot \nabla\right) u^0(x), \tag{2.12}$$

because, for $t = h$, we have, $\xi^h(x, h) = x$.

In the second place let us look at $(v^h \cdot \nabla) v^h$ in $t = h$, for arbitrary $h \in [0, \infty)$ and $t \in [0, \infty)$

$$\left(v^h(x, t) \cdot \nabla\right) v^h(x, t)\bigg|_{t=h} = \left((c + x + u^0(x)) \cdot \nabla\right)(c + x + u^0(x)). \tag{2.13}$$

This leads to

$$\left(v^h(x, t) \cdot \nabla\right) v^h(x, t)\bigg|_{t=h} = (c + x + u^0(x)) + \left((c + x + u^0(x)) \cdot \nabla\right) u^0(x). \tag{2.14}$$

In the third place, in $\mathcal{N}_{v^h}$ we must have

$$\nu \nabla^2 v^h(x, t)\big|_{t=h} = \nu \nabla^2 u^0(x). \tag{2.15}$$

The three results (2.12), (2.14) and (2.15) then give, for $t = h, h \in [0, \infty)$,

$$\mathcal{N}_{v^h}(v^h(x, t))\big|_{t=h} = -\left((x + u^0(x)) \cdot \nabla\right) u^0(x) +$$
$$(c + x + u^0(x)) + \left((c + x + u^0(x)) \cdot \nabla\right) u^0(x) - \nu \nabla^2 u^0(x). \tag{2.16}$$

Or, after rewriting and noting that $\mathcal{N}_{v^h}\left(v^h(x,t)\right)\big|_{t=h} = f(x,h) - (\nabla p)(x,h)$, it follows

$$(c+x) + \mathcal{D}_{c,\nu} u^0(x) = f(x,h) - (\nabla p)(x,h), \tag{2.17}$$

with, $\mathcal{D}_{c,\nu} \equiv 1 + c \cdot \nabla - \nu \nabla^2$. Note,

$$\nabla p = \text{grad}(p) = \left(\frac{\partial p}{\partial x_1}, \frac{\partial p}{\partial x_2}, \frac{\partial p}{\partial x_3}\right).$$

The gradient of the scalar pressure, $p = p(x,t)$ with for arbitrary $p$, gives, $\text{curl grad } p(x,t) = 0$.

### 2.5. Selection of f, requirement (9) (Fefferman, 2000)

If we note that $f$ must be in $\mathcal{G}$, defined in (2.2), then we can select $u^0$ from $\mathcal{U}$, defined in (2.1) such that a stationary $f = f^{crl}$ is given by

$$f^{crl}(x,t) = f^{crl}(x,0) = \nabla \times u^0(x) + \mathcal{D}_{c,\nu} u^0(x) \equiv f^{crl}(x) \tag{2.18}$$

and is in $\mathcal{G}$, see (2.2). Note that the $f^{crl}$ must be divergence free. We have $f^{crl}(x + e_j) = f^{crl}(x)$. Note also that condition (9) of Fefferman (2000) for stationary force, suppressing the notation of the $t = 0$ in the function, is

$$\left|\partial_x^\alpha f^{crl}(x)\right| \leq C_{\alpha,0,K}. \tag{2.19}$$

The constant $K$ is no longer of importance to the upper limit in (2.19). For completeness, the original condition is given in (1.4) and used in the definition in Equation (2.2). If $t = 0$, then substituting this on both sides of the inequality expressing the condition we find is $\left|\partial_x^\alpha f^{crl}(x)\right| \leq C_{\alpha,0}$.

### 2.6. Subset criteria for $u^0$ selection and their use in the breakdown

It is furthermore assumed that:

- $u^0$ is not a vector function of absolute constants,
- $u^0$ is not a gradient vector of an arbitrary proper function $\phi(x)$ and
- $\nabla \times \nabla \times u^0 \not\equiv 0$. This is possible because we have $C^N\left(\mathbb{R}^3\right) u^0$ functions $1 < N \sim$ large.

The previous three points are the constitutive description of a subset of functions of $\mathcal{U}$, given in (2.1). A fourth point is added subsequently. Looking at (2.18) and at (2.19), we may note that the function $u^0$ must be bounded

$$\left|\partial_x^{\alpha_u} u^0(x)\right| \leq C_{\alpha_u} \tag{2.20}$$

in order to satisfy $f^{crl} \in \mathcal{G}$ i.e. its equivalent in (2.19) in combination with the $f^{crl}(x) = \nabla \times u^0(x) + \mathcal{D}_{c,\nu} u^0(x)$ of (2.18). We then denote this series of additional conditions on $u^0$ with $\mathcal{U}' \subset \mathcal{U}$ and select $u^0 \in \mathcal{U}'$. Of course, the definition of $\mathcal{U}$ is such that we are able to have a non empty subset $\mathcal{U}'$. Moreover, because of $\mathcal{D}_{c,\nu} u^0(x + e_j) = \mathcal{D}_{c,\nu} u^0(x)$ and $f \in \mathcal{G}$, it follows from from (2.17) that

$$\nabla p(x + e_j, h) - \nabla p(x, h) = -e_j. \tag{2.21}$$

Hence, if $p$ is possible in the NS equation with $(u^0, f)$, the gradient $p$ cannot be spatial periodic when $v^h$ in (2.4) is not spatial periodic. This agrees with the fact that $v^h$ obeys the type D requirement of not being a solution of Equation (10) of Fefferman (2000).

If we then substitute $f^{crl}$ given in (2.18) into Equation (2.17), noting that we may write $f^{crl}(x) = f(x, h)$, then, because the left hand side of (2.17), i.e. $(c + x) + D_{c,\nu} u^0(x) = f(x, h) - (\nabla p)(x, h)$, does not contain $h$ dependence, it follows that for arbitrary $h \in [0, \infty)$ and $c$ an absolute constant vector,

$$\nabla p(x, h) = \nabla \times u^0(x) - (c + x). \tag{2.22}$$

Hence, it suffices to inspect $p(x, h) = p(x)$. We note that $\nabla \times c = \nabla \times x = 0$. Hence,

$$\nabla \times \nabla p(x) = \nabla \times \nabla \times u^0(x). \tag{2.23}$$

Because of the use of $C^N(\mathbb{R}^3)$, $1 < N \sim$ large, functions, we are allowed to take the curl on left and right hand side of (2.22) and note that because $u^0 \in \mathcal{U}'$, Equation (2.23) cannot be fulfilled. This is so because curl on grad results for all $p$ into the zero vector on the left hand side of this Equation (2.23) while on the right hand side $\nabla \times \nabla \times u^0(x)$ is not the zero vector by necessity. It was assumed that $u^0 \in \mathcal{U}'$, which entails, $\nabla \times \nabla \times u^0(x) \neq 0$.

Hence, with $f^{crl} \in \mathcal{G}$ and $u^0 \in \mathcal{U}'$ it is possible to define a function $v^h$ in (2.4), using a $u^0$ from $\mathcal{U}'$ with arbitrary $h \in [0, \infty)$ such that it is impossible to find a $p = p(x, t)$ function. Hence it is not possible to find $(p, v^h)$ that solves the NS Equation (1.1) with a valid $(u^0, f^{crl})$. This represents requirement (1) of Fefferman (2000).

Also, in anticipation of the discussion, in $d = 2$, the curl is a scalar. See e.g. Aris (1989, p. 223). Hence, the curl geometry in $d = 2$ is quite different from the one in $d = 3$.

### 2.7. Euler equation

From the previous analysis of the NS equation a differential form is obtained with, $\mathcal{D}_{c,\nu} \equiv 1 + c \cdot \nabla - \nu \nabla^2$. In the Euler equation there is no viscosity, i.e. $\nu = 0$, (Papanastasiou et al., 2000). The differential form then is equal to $\mathcal{D}_{c,\nu} \equiv 1 + c \cdot \nabla$. In that particular case we can also select, in a way similar to (2.18), using $\nu = 0$, a $f^{crl}(x)$ force vector function from $\mathcal{G}$. This leads us to the same impossibility as presented in (2.23). The impossibility amounts to, on the one hand the $d = 3$ gradient of the pressure, which vanishes under $d = 3$ curl operation and $\nabla \times \nabla \times u^0(x) \neq 0$ for a $u^0 \in \mathcal{U}'$ on the other hand. The reason for a similar breakdown is that the NS and the Euler equation share the total differential $\left(\frac{Du}{Dt}\right)$ form.

### 2.8. Compound form

The previous analyses of both the NS and the Euler equation can be generalized with the use of the set of intervals $\Lambda$ defined by

$$\Lambda = \{\lambda^h \mid \lambda^h = [h, h + \delta), h \in [0, \infty), \lambda^h \cap \lambda^{h'} = \emptyset, h \neq h'\}. \tag{2.24}$$

The global function $u$ can subsequently be based on $\Lambda$ and is defined by

$$u(x, t) = \lim_{0 < \delta \to 0} \sum_{\lambda \in \Lambda} \chi_\lambda(t) v^{\inf(\lambda)}(x, t). \tag{2.25}$$

Here, $\chi_\lambda(t) = 1$, when, $t \in \lambda$ and $\chi_\lambda(t) = 0$, when, $t \notin \lambda$. Hence, we can see that $u(x, 0) = v^h(x, 0)$ for $h = 0$. The function $v^{\inf(\lambda)}(x, t)$ is defined in (2.4) with $h = \inf(\lambda)$. The definition of the form in (2.25) resembles the use of discreteness in the Feynman path integral, Feynman and Hibbs (1965, pp. 30–39) and Kumano-go (2008).

## 3. Conclusion and discussion

The definition of *u* in (2.25) is an increasingly better -and unstoppable- approximation of the possibility that $u(x,t) = v^h(x,t)$ and *t* in $[h, h+\delta)$. The analysis can be maintained because of the continuum between *h* and $h+\delta$, with $0 < \delta \to 0$. Hence, conclusions for $v^h$ have their impact on *u*. If no $v^h$ can be found then no *u* can be found either. Therefore, it is sufficient to only look at the D breakdown yes-or-no possibility of $v^h$ defined in (2.3) and (2.4).

Furthermore, the curl in $d=3$ is an operation, $\text{curl}:\mathbb{R}^3 \to \mathbb{R}^3$. The difference with $d=2$ is that in that case the **curl** is a scalar (Aris, 1989). Note btw that if one employs the 3-curl to compute the curl of a $d=2$ vector in a space spanned by e.g. $e_1$ and $e_2$, by extension such that e.g. $u_3=0$, then, $\nabla \times u$, projects out of the space spanned by $e_1$ and $e_2$. Hence, such an extension is not a $d=2$ curl.

Nevertheless it can be noted that in a possible exact breakdown type D for $d=2$, home made curl operators can perhaps provide sufficient tooling. An interesting example is based on the home-made outer product, $\times_2$. It is defined for $v=(v_1,v_2)$ and $u=(u_1,u_2)$ by $v \times_2 u = (e_1 v_2 - e_2 v_1)(v_1 u_2 - v_2 u_1)$. With it we define a curl in $d=2$ for $\partial_j = \frac{\partial}{\partial x_j}$ and $\text{grad}_2 = (\partial_1, \partial_2)$, as

$$\text{curl}_2 w = \text{grad}_2 \times_2 w$$

and $w = w(x_1, x_2, t) \in \mathbb{R}^2$. The curl is obtained from our $\times_2$ definition when $v_1 \to \partial_1$ and $v_2 \to \partial_2$ is substituted. Note the required $\text{div}_2 \text{ curl}_2 w = 0$ and $\text{curl}_2 \text{ grad}_2 w = 0$. Furthermore, the operation $\times_2$ is right-hand linear, i.e. $w \times_2 (u+v) = w \times_2 u + w \times_2 v$. The $\times_2$ is however not left-hand linear, $(u+v) \times_2 w \neq u \times_2 w + v \times_2 w$. Another point that can be raised against $\text{curl}_2 = \text{grad}_2 \times_2$ is that for parity transformation $(\partial_1, \partial_2) \to (-\partial_1, -\partial_2)$, by definition we must have $(-\text{grad}_2) \times_2 = \text{grad}_2 \times_2$. Consequently, the $d=2$ breakdown equation, $\text{grad}_2 p(x) = \text{grad}_2 \times_2 u^0(x) - (c+x)$ shows an inconsistency between $p(-x) = sp(x)$ and $u^0(-x) = su^0(x)$, $s = \pm 1$, on the one hand and $\text{curl}_2 \text{ curl}_2 u^0(x) \not\equiv 0$ on the other hand. This awkward behavior is no big surprise because an official $\text{curl}_2 : \mathbb{R}^2 \to \mathbb{R}^2$ is non-existent.

To return to our main conclusion. There is a type D breakdown in $d=3$ and it does not disagree with the prove (Ladyzhenskaya, 1969) that there is a generalized solution in $d=2$ space. This is a valid statement despite the possibility of a most likely fit-for-purpose home made operator $\text{curl}_2$.

Interestingly, the qualitative difference between $d=2$ and $d=3$ spatial dimensions occurs elsewhere too. There is a remarkable difference in behavior, also observed experimentally (Laughlin, 1983), of quantum spin statistics in $d=2$ versus $d=3$. This difference is between complete freedom of spin statistics of 'particles' in a $d=2$ space, i.e. the so called anyon spin, opposed to the spin statistics of particles that are allowed the freedom of a $d=3$ space. We see a discrete number (e.g. 2 for electrons) of possibilities for spin states in $d=3$. The research into this difference was initiated by Leinaas and Myrheim (1977) and Wilckzek (1982) and further explained and deepened by e.g. Lerda (1992) and Kitaev (2006). Apparently adding or subtracting a spatial degree of freedom provides significant changes of behavior of a system. This apparently goes for systems under the rule of quantum mechanics as well as for systems under the rule of continuum mechanics.

Concerning the $C^N$ set of functions, the question is raised whether a possible rejection of the breakdown presented in this paper can be based on its use of $1 < N \sim \text{large}$. The reader is invited to capture, with genuine differentiation operations, the difference between *N* very large and, growing if necessary but finite, as opposed to infinite continuous differentiable. The set of functions, $\mathcal{U}'$, probably also has an untraceable large cardinality. The breakdown of $(p, u)$ with $C^N$, with $N \to \infty$ requirement can be replaced with *N* 'untraceable large' and, growing when considered necessary. In the final analysis, the NS problem with its requirements is rooted in the physics of fluids. Note however that the present analysis also does not escape completely from the illusion: "infinity is known to us, it is just very large". E.g. use is made in (2.25) of the continuum between two distinct points on the real axis. However, if the restriction to untraceable large but countable is employed in the context of sets of functions, i.e. to $C^N(\mathbb{R}^3)$ functions with $N \in \mathbb{N}$ and $1 < N \sim \text{large}$, then the continuum assumption of the real axis appears valid in the limit. This part of the discussion shows traits of constructive analysis versus classical foundations of mathematics. See e.g. Bishop (1967) or finitistic mathematical foundations (Ye, 2011). Note that in both cases a similar definition of $\mathbb{R}$, see e.g. Ye (2011, p. 73) and chapter 1 of Bishop (1967), is used. The present author remains on the practical side of the debate. For practical purposes, $\infty$, is not a known number.

To wrap it up. In this paper a $d = 3$, D-type of breakdown (Fefferman, 2000) was demonstrated for the NS equation. For valid $(u^0, f)$ there is no $(p, u)$ possible that exactly solves the NS.

### Acknowledgements
The author would like to thank the editor and the reviewers. The author also wants to thank F. Kaptein for reading the manuscript.

### Funding
The author received no direct funding for this research.

### Author details
Han Geurdes[1]
E-mail: han.geurdes@gmail.com
ORCID ID: http://orcid.org/0000-0002-7487-1875
[1] Geurdes consultancy, C vd Lijnstraat 164, 2593 NN Den Haag, Netherlands.

### Citation information
Cite this article as: A type D breakdown of the Navier Stokes equation in $d = 3$ spatial dimensions, Han Geurdes, *Cogent Mathematics* (2017), 4: 1284293.